\documentclass[pdflatex,sn-mathphys-num]{sn-jnl}

\def \sla{\slash {\!\!\!}}

\usepackage{graphicx}%
\usepackage{multirow}%
\usepackage{amsmath,amssymb,amsfonts}%
\usepackage{amsthm}%
\usepackage{mathrsfs}%
\usepackage[title]{appendix}%
\usepackage{xcolor}%
\usepackage{textcomp}%
\usepackage{manyfoot}%
\usepackage{booktabs}%
\usepackage{algorithm}%
\usepackage{algorithmicx}%
\usepackage{algpseudocode}%
\usepackage{listings}%

\theoremstyle{thmstyleone}%
\theoremstyle{thmstyletwo}%

\theoremstyle{thmstylethree}%

\begin{document}

\title[Spontaneous Symmetry Breaking: ]{ Massive gauge boson  and  Higgs boson as the vestiges of non-Fock vacuum}


\author*[1]{\fnm{Shun-ichiro} \sur{Koh}}\email{koh@kochi-u.ac.jp}



\affil*[1]{\orgdiv{Divison of Physics}, \orgname{Kochi University}, \orgaddress{\street{2-5-1, Akebono-cho}, \city{Kochi}, \postcode{780}, \state{Kochi}, \country{Japan}}}




\abstract{A microscopic model of the Brout-Englert-Higgs (BEH) mechanism is proposed.  Massless fermions and antifermions  do not  belong  to the  Fock space with definite particle-number distribution, but belong  to a non-Fock  space with  indefinite  one.  From this vast  space,    their  ground-state  is selected by a  kinematical condition.   Due to the interaction   between   them,  the Fock state in which  fermions and antifermions are massive is restored, but this state has the vestiges of the non-Fock state in the massive gauge boson and the Higgs boson.   In the non-Fock state, massless fermions and antifermions  exist  as  pairs,   and   behave as  quasi bosons within a small space-time region.  Due to Bose statistics, the direction of  their  center-of-mass  motion is parallel to each other, and their   transverse excitations are suppressed by an energy gap, making the gauge boson coupled to them massive.   The Higgs boson is  not  an elementary particle,  but  a quasiparticle appearing after the Fock vacuum is restored.  }




\maketitle

\section{Introduction}\label{sec1}
Spontaneous breaking of  symmetry in vacuum  is divided roughly  into two categories.  The first is  the Brout-Englert-Higgs (BEH) mechanism  \cite{eng}\cite{hig}, which is an extension of the Stueckelberg formalism \cite{stu}.   The BEH mechanism   is a simple model in which both of symmetry breaking and its  consequence are derived by adding to $-\frac{1}{4}F^{\mu\nu}F_{\mu\nu}$  the following  gauge coupling and Higgs potential  of a scalar field $h$,    
\begin{equation}
    L_h(x)= |(i\partial_{\mu}+gB_{\mu})h |^2 - \mu ^2|h|^2 -\lambda |h|^4  .
		 \label{eq:01}
\end{equation}
 Switching the sign of $\mu ^2$ leads to the broken-symmetry vacuum with the vacuum condensate $v_h$. This $v_h$ gives a mass  $m_B=gv_h$ to  the vector-Abelian-gauge field $B_{\mu}$.  The above  $h$ becomes $v_h+ h_1+ih_2$  composed of the Higgs field $h_1$ and the Goldstone mode $h_2$. (Stueckelberg  took only the Goldstone mode, and proposed a simple gauge-invariant model of massive gauge boson. )
 
For the electroweak interaction in which the weak interaction is effective only in the left-handed fermions,  the Glashaw-Weinberg-Salam (GWS) model  \cite{gla}\cite{wei}\cite{sal} used  the BEH mechanism for the chiral-symmetric  Lagrangian density  $-\frac{1}{4}F^{a\mu\nu}F^a_{\mu\nu}+\bar{\varphi}(i\partial_{\mu}+ igA^a_{\mu}\tau_a+ig'B_{\mu})\gamma^{\mu}\varphi $,  and  gave a finite mass $m_f$ to massless fermion $\varphi $ by adding the Yukawa interaction as follows
\begin{equation}
    L'_h(x)= L_h(x)    - \frac{m_f}{v_h} h \bar{\varphi} \varphi  .
		 \label{eq:011}
\end{equation}
This model  predicted the  W and Z bosons and the  Higgs  boson found in 1983 \cite{ua1}\cite{ua2} and in   2012   \cite{atl}\cite{cms},  respectively,  and  does not contradict   almost all experimental results  to date.   The Higgs potential  explains many phenomena using  a small number of parameters. This is  because it  plays  a double role:  the role of causing symmetry breaking in vacuum and that of  predicting    the mass of Higgs boson.  Furthermore,   it  stabilizes the broken-symmetry vacuum, and it  represents  the interaction between the Higgs bosons. In this sense,  the Higgs potential is a simple and   economical  model, and  it also has the flexibility to adapt  to  complex situations of the eletroweak interaction.   For its simplicity,  however, we face  some difficult problems.  The switch of the sign of $\mu ^2$ is an ad hoc assumption without explanation.  The origin of the vacuum condensate $v_h$ is not clear.   When the Higgs potential  is used in the perturbation calculation,  we must care a lot about the intricate cancellation of the quadratic divergence.

The second  is  dynamical symmetry  breaking \cite{nam}\cite{jac}\cite{cor}.  For a broken   symmetry  in our world,  we can consider  a  symmetric world.  Although all material particles have their own masses,   massless material particles  are considered.  Beginning with the Fock vacuum, the  interaction  through gauge field  makes this vacuum unstable, and forms a new vacuum, which belongs to a non-Fock representation at the infinite-volume limit.  This  is considered as  a physical process which we can really  observe in experiments. Phase transitions in the non-relativistic condensed matter  belong to this category. To understand the mechanism behind the BEH mechanism, many attempts have been made using an analogy with them.   This is a probable analogy,  but to our knowledge,   we do not yet  clearly understand  how such a symmetry breaking occurs in vacuum.

 In order to understand the physics behind,   we  look at the problem from a  different angle.
Unlike massive fermions,    massless fermions  in the Fock state   are  questionable, because finite  mass, which  supports the Fock state from behind,  is not there. 
So long as we begin with a chiral-symmetric Lagrangian, the  non-Fock state is needed from the beginning, and it is not necessary to derive it  dynamically from the Fock state. The order of  this argument  may be the reverse of the order of the  logic  in dynamical symmetry breaking. Kinematics of massless  antifermion rather than dynamics determines the non-Fock state. 
There massless fermions acquire a mass by their  interaction, and the  Fock state is restored, but this state is different from the conventional Fock state in that  the massive gauge boson and the Higgs boson exist as  the vestiges of the non-Fock state.  In  this non-Fock state, massless fermion-antifermion pairs respond to the gauge field as quasi bosons, which  causes massive gauge boson.  It is experimentally confirmed that the fermion's  coupling to Higgs boson is proportional to fermion's mass. If we take a step  of approximation further  in  the mass acquisition of fermion, the Higgs-like boson  will appear.  From  this viewpoint,   the Higgs boson   is  not an elementary particle, but a quasiparticle appearing after the Fock vacuum is restored.

This paper is organized as follows.  
 In  Section 2, we reconsider massless fermion in the non-Fock  state.    For  its  ground state,  a  kinematical condition  is used  as a route  to the  non-Fock vacuum.  Whereas  the Nambu Jona-Lasinio state is recognized   as such a vacuum,   the massless fermion is an unstable existence by nature  and  becomes massive  whatever   interaction acts on them.    In Section 3,  the  quasi-boson's aspect of  massless   fermion-antifermion  pair is explained.   To describe them,   some  constants  of   vacuum are needed, and the $v_h$, $\mu$ and $\lambda$ in  Eq.(\ref{eq:01}) are examples of such constants. The question  is how to introduce these constants  in a physically natural way.  We define microscopic  three constants of the non-Fock  vacuum in position space.  In Section.4, with these  constants and quantum statistics, we  propose an origin of the vacuum condensate $v_h$, and calculate the gauge-boson's mass.    In Section 5,  we  calculate Higgs mass $m_H$ as an excitation energy of  a collective motion of massive fermion and antifermion mediated by massive gauge boson.   Parameters in  the Higgs potential are reproduced  by these constants. In Section 6, we discuss some  implications   for  the BEH mechanism. 
 
 \section {Non-Fock state under  chiral-symmetric Lagrangian   }

 Let us examine  the   vacuum under the   simple chiral-symmetric  Lagrangian density,    
\begin{equation}
		L_0(x)= -\frac{1}{4}F^{\mu\nu}F_{\mu\nu}+\bar{\varphi}(i\partial_{\mu}+gB_{\mu})\gamma^{\mu}\varphi   ,
		 \label{eq:1}
\end{equation} 
where $B_{\mu}$ is a U(1) gauge field, and   $\varphi$ is a massless Dirac  fermi field \cite{massless},
\begin{equation}
		\varphi(x)= \frac{1}{\sqrt{V}}\sum_{p,s} 
		     \left[ a^s(\mbox{\boldmath $p$})u^s(p)e^{-ipx}+ b^{s\dagger }(\mbox{\boldmath $p$})v^s(p)e^{ipx} \right]  .
             		       \label{eq:211} 
\end{equation}
\subsection{Massless fermion  in the non-Fock  state }
  In the case of  massive fermions,   a state is divided  into subspaces with definite numbers  of particles,  the existence of which  is supported  by a  finite mass.   Fock state  is a linear combination of basic vectors  $\sum_n c_n |n_1,n_2, \cdots  \rangle$ where $n=(n_1,n_2 \cdots )$ is a particle-number distribution.  Whereas the degree of freedom $\lambda$ in  $n_{\lambda}$  extends to  infinity,    only finite number of $n_{\lambda}$ is nonzero in  $ |n_1,n_2, \cdots  \rangle$ of the Fock state, such as  $\sum_n |c_n|^2 <\infty$, and the sequence $n_1n_2 \cdots$ represents a decimal.   As a result, the   number of  basic vectors is   countable.  The operators  $ a^s(\mbox{\boldmath $p$})$ and $ b^{s\dagger }(\mbox{\boldmath $p$})$, which strictly specifies the number of particle as a lowering or  raising   operator,  is closely linked to   Fock state.   For  the massless fermion, however,  there is no reason for  this rule to hold,   because mass is no longer an indicator of  individual particle, and  momentum remains as a continuous parameter.  To describe this parameter,  nonzero  sequence  $n_1n_2 \cdots$   extending  to infinity should be prepared,   and  the number of basic vectors  is  uncountable  as in  the cardinality of  real numbers.  Such a state is not divided  sharply  into  subspaces as in the Fock state \cite{flato}. The  vacuum of massless fermion and antifermion  is a non-Fock  state with indefinite  particle-number. 
   
 This indefinite  particle-number is  realized by the dynamics  in $L_0(x)$   as follows.  If we consider massless particles to be real entities,  they become a  subject of   the theory of relativity, even if their velocity  is not so large as that of light.  Due  to pair production from $B_{\mu}$,   the number of   massless fermion and antifermion  is not fixed, because  mass no longer gives  a threshold energy in the pair production. The state  with indefinite  particle-number is preferable.  
  The space expanded by uncountable  basic vectors, however,   is a vast state-space. Hence,  we should proceed  with the help of kinematical condition,  because kinematics does not depend on the details of the interaction,  and kinematics is more fundamental than dynamics in this case. For massless particle,  a new raising and lowering  operator reflecting this indefinite  particle-number   is needed. 

\subsection{Kinematical condition}
 
  For the gauge field, the moving massless  fermions  or antifermions are nothing but charge's and spin's  vehicles.
   Since vacuum is neutral in charge and isotropic in space, the  annihilation  of negative charge in  $a^s(\mbox{\boldmath $p$})$  at  the departure of  massless fermion brings about the same effect to the  gauge field $B_{\mu}(\mbox{\boldmath $p$})$ 
 as    the  creation  of positive charge in $b^{s\dagger }(-\mbox{\boldmath $p$})$ at  the arrival of   massless antifermion  with opposite momentum and spin. 
   Furthermore, momentum, electric charge and spin, which prescribe all  properties of massless fermion,  are not positive definite, and therefore   the effect of $a^s(\mbox{\boldmath $p$})$ and that of  $b^{s\dagger }(-\mbox{\boldmath $p$})$ on the state are  equivalent.   {\it There is no reason to assign the  equivalent  operations  on the state  to two different operators.}  For   this   to be ensured,   a new raising and lowering  operator $\widetilde{a}^s(\mbox{\boldmath $p$})$       is defined as   follows
\begin{equation}
		\widetilde{a}^s(\mbox{\boldmath $p$})= \cos \theta_{\mbox{\boldmath $p$}} a^s(\mbox{\boldmath $p$})
		                                         + \sin \theta_{\mbox{\boldmath $p$}}b^{s\dagger }(-\mbox{\boldmath $p$})   .
				                	 \label{eq:8}
\end{equation}
If the fermion is massive,  the annihilation of fermion and the  creation of antifermion  cause different effects  on the state   because mass is positive definite, and such a raising and lowering  operator does not exist.   Hence, it is  a characteristic of  the massless fermion and antifermion.  

 The necessity of antifermion in   Eq.(\ref{eq:8}) changes according to the relative momentum   between the  particle and the observer, which is reflected in $\sin \theta_{\mbox{\boldmath $p$}} $.  When $\mbox{\boldmath $p$} =0$,  the difference   between $ a^s(\mbox{\boldmath $p$})$ and $b^{s\dagger }(-\mbox{\boldmath $p$})$ in  momentum vanishes. Therefore,     the  operators  $a^s(\mbox{\boldmath $p$})$ and $b^{s\dagger }(-\mbox{\boldmath $p$}) $  have  the same importance  for the   observer,   resulting in  $\cos \theta_{\mbox{\boldmath $p$}}=\sin \theta_{\mbox{\boldmath $p$}}$ at $\mbox{\boldmath $p$} =0$.  On the contrary, when $\mbox{\boldmath $p$}^2 \rightarrow \infty$,   such  observers  cannot  be found, and antifermion is not needed. Hence,   $\sin \theta_{\mbox{\boldmath $p$}}  \rightarrow 0$  is expected.  The situation is same  also for the  antifermion.   A  new  operator $\widetilde{b}^s(-\mbox{\boldmath $p$})$ is defined  as  a superposition of  the annihilation  of massless antifermion  and the creation  of massless fermion   
\begin{equation}
\widetilde{b}^s(-\mbox{\boldmath $p$})= \cos \theta_{\mbox{\boldmath $p$}} b^s(-\mbox{\boldmath $p$})
		                                         - \sin \theta_{\mbox{\boldmath $p$}}a^{s\dagger }(\mbox{\boldmath $p$})   .
				                	 \label{eq:9}
\end{equation}
 This $\widetilde{b}^s(-\mbox{\boldmath $p$})$ is orthogonal to $\widetilde{a}^s(-\mbox{\boldmath $p$})$.
 
  The representation of non-Fock world  is defined by  its vacuum. Really,   a  vacuum  $|\widetilde{0}\rangle$ satisfying   $\widetilde{a}^s(\mbox{\boldmath $p$})|\widetilde{0}\rangle =\widetilde{b}^s(-\mbox{\boldmath $p$})|\widetilde{0}\rangle=0$ exists.   The explicit form of  $|\widetilde{0}\rangle$ is inferred as follows.  For  massless fermions  observed that they  move as fast as light,  relative momentum  between two different observers  has no meaning in practice, so that $ \cos \theta_{\mbox{\boldmath $p$}}  \rightarrow 1$ is required  in  Eqs.(\ref{eq:8}) and (\ref{eq:9}),  and therefore   $|\widetilde{0}\rangle$ should  include $ \cos \theta_{\mbox{\boldmath $p$}} |0\rangle$ for this limit.  Conversely,  for  fermions  observed that they move  with   small momentum,     not only massless  fermions with $\mbox{\boldmath $p$}$ but also  antifermions with $-\mbox{\boldmath $p$}$ is necessary,  and  they  should coexist  in $|\widetilde{0}\rangle$  as  $b^{s\dagger}(-\mbox{\boldmath $p$}) a^{s\dagger}(\mbox{\boldmath $p$}) |0\rangle$.  
 The simplest possible form of $|\widetilde{0}\rangle$ including these two  cases    is a  superpositions of $ \cos \theta_{\mbox{\boldmath $p$}} |0\rangle$ and  $\sin \theta_{\mbox{\boldmath $p$}}b^{s\dagger}(-\mbox{\boldmath $p$}) a^{s\dagger}(\mbox{\boldmath $p$}) |0\rangle$. 
  The vacuum $|\widetilde{0}\rangle$  is a direct product of such superpositions  for all $\mbox{\boldmath $p$}$  (see Appendix.A) 
\begin{equation}
		|\widetilde{0}\rangle=\prod_{p,s} \left[ \cos \theta_{\mbox{\boldmath $p$}} 
    +\sin \theta_{\mbox{\boldmath $p$}} e^{i\alpha(x)}b^{s\dagger}(-\mbox{\boldmath $p$}) a^{s\dagger}(\mbox{\boldmath $p$}) \right]|0 \rangle   .
						                	 \label{eq:16}
\end{equation}
This  $|\widetilde{0}\rangle$ is an example of non-Fock states in which the number of basic vector is uncountable as represented by a continuous weight-factor  $\sin \theta_{\mbox{\boldmath $p$}} $, which is selected  by the kinematical condition  among various candidates in the vast state-space. 

 This  $|\widetilde{0}\rangle$ was first introduced to elementary-particle physics by \cite{nam} in analogy with superconductivity \cite{sup}.  However, it is not a form specific  to  this analogy,  but the first approximation of  the  general form  for the   relativistic massless many-body state \cite{normal}.  
 The analogy with  superconductivity has the  following limits.  In metals, the momentum of  electrons   represents  the relative motion of electrons   to the center-of-mass of  crystal.  In the physical vacuum, the naive   analogy such as  the center-of-mass of world cannot be carried over, because the significant momentum  is only  that  of  relative motion between  observer and  particle. The derivation  via  Eqs.(\ref{eq:8}) and (\ref{eq:9}), which uses the  relative momentum between them,  is a natural way to describe physical  vacuum  \cite{bogo}.
 
 \subsection {Restoration of Fock vacuum}
  The kinematical condition, rather than dynamics,  leads to the non-Fock vacuum $|\widetilde{0}\rangle$.  However, particles  in  interacting with others  cannot exist as  massless ones   unless they are protected   by some stronger principle. 
  The dynamics  in $L_0(x)$ this time  prevents the exact realization of  such a non-Fock vacuum.    The  restoration  of the Fock vacuum occurs  in the   following way.

  Massless fermions and antifermions   interact to each other  via $B_{\mu}$.   The  interaction between them, regardless of being  repulsive or attractive,  induces the motion of other massless  fermions and antifermion. 
 As a result, the motion of original   fermions and antifermion is not easily affected by the external force, which  comes back to  the massless  fermions and antifermion  as an inertial  mass. 
   The  simplest way to demonstrate this mechanism  is a mean-field approximation.  The mean field is an approximation that neglects the long-range fluctuation.   As long as  the gauge boson is  massless,  the interaction between fermions is a long-range one,  which induces  the long-range  longitudinal excitation. Such excitations  involve a large amount of particles and  require a high excitation-energy (like the plasmon in the electron gas). Hence,  the long-range fluctuation is suppressed, and the mean field  is an useful way.  Furthermore, the restoration of the Fock vacuum is related to the stability of the  world. 
   For this world to  continue to exist  forever,  such a process should depend, not on the time-dependent dynamics, but on the time-independent one that is well described by the mean field  \cite{vec}. 
 Assuming  a  scalar mean-field $U_0$,  we consider
\begin{equation}
		\bar{\varphi}(x)(  i \sla{\partial }+ U_0) \varphi (x).
				                	 \label{eq:751}
\end{equation}
Following the  procedure  in  \cite{sup},    the  condition that Eq.(\ref{eq:751}) is  diagonal with respect to 
$ \widetilde{a}^{s}(\mbox{\boldmath $p$})$ and $\widetilde{b}^{s}(\mbox{\boldmath $p$})$ is as follows 
\begin{equation}
	\cos ^2\theta_{\mbox{\boldmath $p$}} =\frac{1}{2}\left(1+\frac{\epsilon_p}{\sqrt{\epsilon_p^2+ U_0^2} }\right)   ,  \quad
	\sin ^2\theta_{\mbox{\boldmath $p$}} =\frac{1}{2}\left(1-\frac{\epsilon_p}{\sqrt{\epsilon_p^2+ U_0^2} }\right)   ,
						                	 \label{eq:23}
\end{equation}
which satisfies the kinematical  requirement  $\cos \theta_{\mbox{\boldmath $p$}}=\sin \theta_{\mbox{\boldmath $p$}}$ at $\mbox{\boldmath $p$}=0$,   and $\sin \theta_{\mbox{\boldmath $p$}} \rightarrow 0$ at $\mbox{\boldmath $p$} \rightarrow \infty$ that  we imposed   in   Eqs.(\ref{eq:8}) and (\ref{eq:9}).  
 The diagonalized form of  Eq.(\ref{eq:751})  includes $\sqrt{\epsilon_p^2+ U_0^2}
[\widetilde{a}^{s\dagger }(\mbox{\boldmath $p$}) \widetilde{a}^{s}(\mbox{\boldmath $p$}) + \widetilde{b}^{s\dagger }(\mbox{\boldmath $p$}) \widetilde{b}^{s}(\mbox{\boldmath $p$})]$, in which  $U_0^2$ is  a  square of the mass $m_f^2$  of the fermion and antifermion. Here, the sign of $U_0$ does not affect the result, hence it does not depend on whether $U_0$ in 
Eq.(\ref{eq:751}) is  attractive or repulsive.   If we  regard  this symmetry breaking  as a direct  analogue of superconductivity, such as the technicolor model,  the attractive interaction is a necessary assumption  to derive  the non-Fock vacuum. However, because the non-Fock vacuum is a logical consequence of the zero mass, such an assumption is not necessary.  

The restored Fock vacuum is different from the conventional Fock vacuum of the massive fermion in that the vestige of the non-Fock vacuum remains in the massive gauge boson and the Higgs boson.

 (a)  The  massless  fermion and antifermion  form  a pair  in $ |\widetilde{0}\rangle$, and these pairs obey  Bose statistics within a  small space-time region after pair production.   Due to Bose statistics, an energy gap appears in the transverse excitation of these pairs, which  makes  this state  robust against   transverse  disturbance, and  makes  the gauge boson massive when it couples to these pairs.  This  will  give us an origin of the vacuum condensate $v_h$ in the Higgs potential, as will be explained in Section 3 and 4.
 
 (b)  When we proceed to the  beyond-mean-field approximation in Eq.(\ref{eq:751}), the Higgs particle appears,  as will be explained in Section 5.

\section{Vacuum condensate}
 \subsection{Massless  fermion-antifermion pairs as quasi bosons }
In the BEH model, the vacuum condensate $v_h$ in $L_h(x)$ implies that vacuum is robust against perturbation. This robustness should be physically explained by the  massless  fermion and antifermion in  Eq.(\ref{eq:16}). 
 When they appear or  disappear  in $ |\widetilde{0}\rangle$,  it is always as a pair   with mutually opposite momenta and spins. For  short  periods  after pair  production, their fields should   overlap   in position space,  and 
  their  behavior is  represented   by  dimensionless  composite  operators  $P_{\mbox{\boldmath $k$}}\equiv b (-\mbox{\boldmath $k$},\downarrow) a(\mbox{\boldmath $k$},\uparrow)$ and  $P^{\dagger}_{\mbox{\boldmath $k$}}\equiv a^{\dagger}(\mbox{\boldmath $k$},\uparrow) b^{\dagger}(-\mbox{\boldmath $k$},\downarrow)$. ($\uparrow, \downarrow$ denote spins.)  Owing to the anti-commutation relation,  the composite  operator $P_{\mbox{\boldmath $k$}}$  has the following equal-time   commutation relation at 
$\mbox{\boldmath $k$}\ne \mbox{\boldmath $k$}'$
  \begin{equation}
  [P_{\mbox{\boldmath $k$}},  P^{\dagger}_{\mbox{\boldmath $k$}'} ]=0
                                         \quad  \mbox{for} \quad \mbox{\boldmath $k$}\ne \mbox{\boldmath $k$}'  ,
		  \label{eq:5.14}
\end{equation}
implying that massless  fermions-antifermions pairs with different momentum follow   Bose statistics. However  when $\mbox{\boldmath $k$}= \mbox{\boldmath $k$}'$, due to  Pauli principle, $P_{\mbox{\boldmath $k$}}$ shows the  following   commutation relation
  \begin{equation}
  [P_{\mbox{\boldmath $k$}}, P^{\dagger}_{\mbox{\boldmath $k$}} ]
              =1-(n_{\mbox{\boldmath $k$},\uparrow}+n_{-\mbox{\boldmath $k$},\downarrow}) , 
                                                    \quad  P_{\mbox{\boldmath $k$}}^2=P^{\dagger 2}_{\mbox{\boldmath $k$}}=0  ,
                         \label{eq:5.15}
\end{equation}
where $n_{\mbox{\boldmath $k$},\uparrow}=a^{\dagger}(\mbox{\boldmath $k$},\uparrow)a(\mbox{\boldmath $k$},\uparrow)$ and  $n_{-\mbox{\boldmath $k$},\downarrow} =b^{\dagger}(-\mbox{\boldmath $k$},\downarrow)b(-\mbox{\boldmath $k$},\downarrow)$. This is a hybrid of boson's one
 $ P_{\mbox{\boldmath $k$}} P^{\dagger}_{\mbox{\boldmath $k$}}=N_{\mbox{\boldmath $k$}}+1$ and fermion's one $ P_{\mbox{\boldmath $k$}} P^{\dagger}_{\mbox{\boldmath $k$}}=1-(n_{\mbox{\boldmath $k$},\uparrow}+n_{-\mbox{\boldmath $k$},\downarrow})$. 
These  commutation relations show that the center-of-mass motion of the pair denoted by $P_{\mbox{\boldmath $k$}}$ and its response to $B_{\mu}$  are  that of  a quasi boson.   

\subsection{ Relativistic quantum field in position space}
\begin{figure}
	 	\centering
\includegraphics [bb=0 0 668 387, scale=0.585] {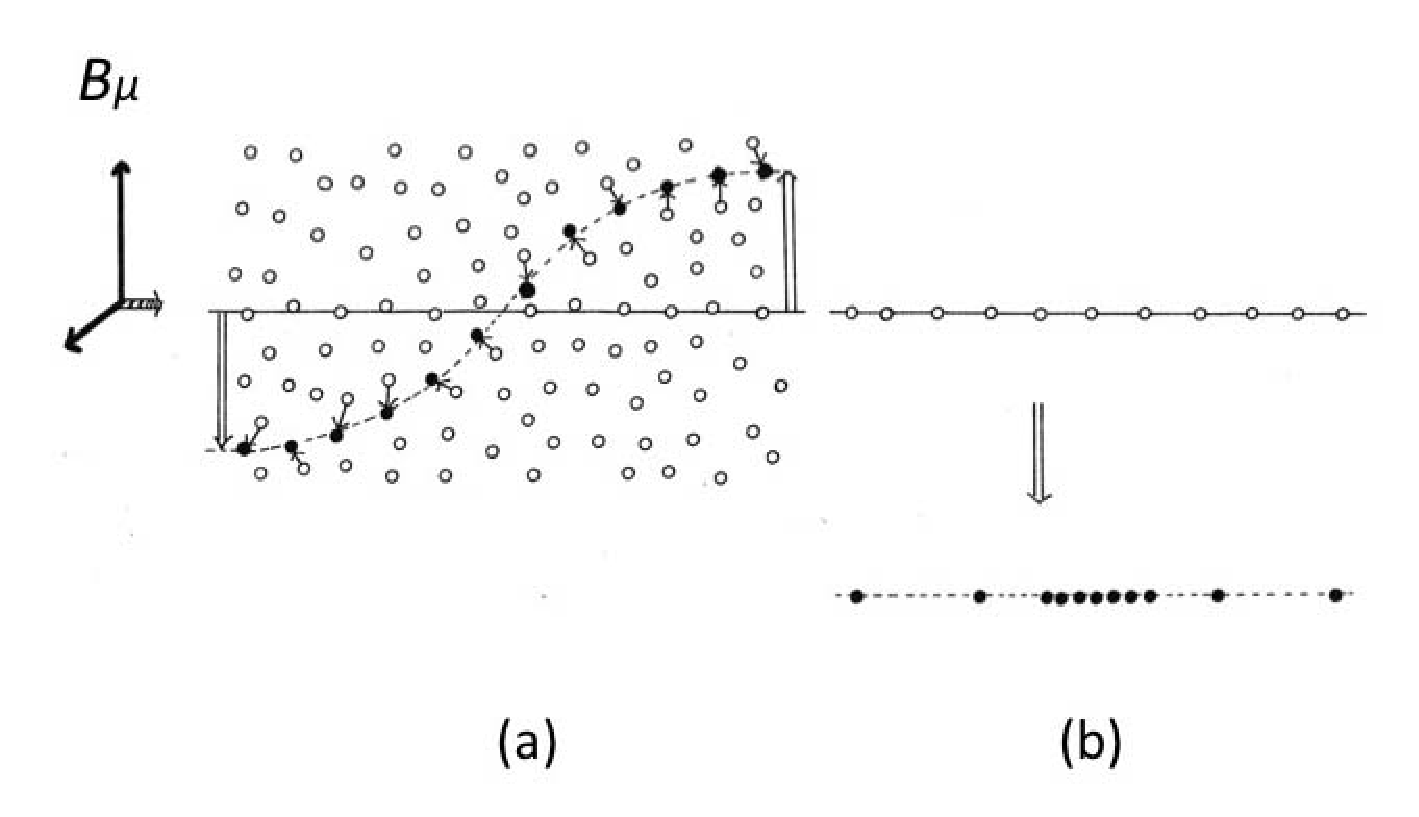}
\caption{A schematic view of the physical vacuum at a  moment  in position space.  Each white  circle blurred with  $\delta x =\hbar c/\epsilon$ represents  a quasi-boson, which  is distributed with a mean distance $d_m$.   (a) The transverse displacements ($dr$) of quasi bosons  from  the horizontal  line ($z$ axis) are coupled to the  gauge field $B_{\mu}$ propagating  along the $z$ axis,  and  (b) the longitudinal displacements ($dz$) are illustrated.    }
\label{fig.311}
\end{figure}

In relativistic quantum field theory, position is complementary not only to momentum, but also to particle number. If we make a precise measurement of position,  it causes  a wide  spread of momentum, and owing to subsequent  pair production, the number of particles becomes  indefinite. However, position of particle is not a completely meaningless concept. In the measurement of position of moving particle with an energy $\epsilon$, the time duration necessary for measurement  using light
$\delta t \simeq \Delta x/c$ satisfies $\delta t >\hbar/c\Delta p>\hbar /cp \geq  \hbar /\epsilon$. Since the particle may move 
$c\delta t > \hbar c /\epsilon$ during this interval,  the least possible  error  of position $\delta x$  is $\hbar c/\epsilon$ \cite{land}. Unless we attempt to specify the position of particle more precisely than $\delta x$,  it  has a physical meaning.  Quasi bosons  appear and  disappear randomly  in position space, but  we consider  for simplicity a {\bf mean distance between quasi bosons} $d_m$, which  is larger than $\delta x$.   Figure \ref{fig.311}  schematically illustrates  the distribution of  these quasi bosons  in position space at a  moment, using  small white  circles being blurred with the uncertainty $\delta x =\hbar c/\epsilon$. The gauge field  $B_{\mu}$  propagates  along the horizontal line.  Since quasi bosons made of fermion and antifermion with different momentum prevail in vacuum, they follow Bose statistics as in  Eq.(\ref{eq:5.14}).   However, since these  quasi bosons are not stable,  their  behavior as  bosons is a limited one in the temporal and spatial sense.   Here we define  a {\bf coherence length} $l_c$ in the  vacuum such as $\delta x <d_m<l_c$ so that  as long as  the spatial  distance between  these quasi bosons  is less than  $l_c$,  they obey Bose statistics.   

\subsection{Statistical gap}
  Consider quasi bosons using   a   composite scalar  field $ f(x)$  made of  $\varphi(x)$ in Eq.(\ref{eq:211})
  \begin{equation}
f(x )= \frac{1}{\sqrt[3]{V}} \sum_{k,s} 
                              [P_{\mbox{\boldmath $k$}} \bar{v}^s(- k)  u^s( k )
                   + P^{\dagger}_{-\mbox{\boldmath $k$}} \bar{u}^s( -k) v^s( k)] e^{ikx}  . 
  \label{eq:2691}
\end{equation}
 (The normalization volume $V$ in $\varphi (x)$ and $f(x)$  is $d_m^3$.)  This $f(x )$ describes the excitations of  bosons  within the coherence length, which  has  the  following  kinetic energy
\begin{equation}
H_{ef}= \int   \left| \frac{\partial }{\partial x_i} f(x) \right|^2   d^3x = - \int   f^{\dagger} (x)  \Delta  f (x)  d^3x .
  \label{eq:2692}
\end{equation}
 In   Figure \ref{fig.311}(a),  quasi bosons are  excited by  the gauge field $B_{\mu}$  from the solid horizontal line to  a dotted wavy line.  Let us consider  their  kinetic energy.   Bose statistics requires   permutation symmetry  for  quasi bosons, and therefore displacements pointed by  a long white arrow from the horizontal line  to black circles  in  Figure \ref{fig.311}(a) is indistinguishable from that by  small arrows between a  black circle  and a neighboring  white circle.    Because the  transverse displacements are perpendicular to the wave vector, they are not additive,  and there are many short displacements  in the excitation. The long transverse  displacements  can be  replaced by the many  short ones.  (In contrast, since the longitudinal displacement is additive as shown  in Figure \ref{fig.311}(b),  the long displacements are dominant,  and the short displacements are exceptional.) The short distance in the denominator of Laplacian operator $ \Delta $ in  Eq.(\ref{eq:2692}) creates a high excitation energy, and an energy gap $\epsilon _0$ from zero appears in the transverse  excitation spectrum.  This energy gap  comes only from statistical property, and we call it  {\bf statistical gap} \cite{fey2}.  The  long  movements  are substantially forbidden  in the {\bf transverse}  motion  of quasi bosons.

This situation is expressed using   a metric tensor $g_{ij}(x)$ in differential geometry. For the transverse and longitudinal distances $r$ and $z$ respectively  in  Figure \ref{fig.311},  the square  of the line element is expressed  as $dl^2=g_{rr}(r)dr^2+ dz^2$.   When the quasi bosons  in  Figure \ref{fig.311}(a)  move in  the transverse direction  within  the coherent spatial  region with a size $l_c$,  a small distance $dr$ such as $d_m$  causes the same effect to $dl^2$ as a large distance $r$.   A simple metric representing this  situation   is given by
\begin{eqnarray}
        g_{rr}(r) &=& \frac{r^2}{d_m^2} , \quad   (d_m \leq r <l_c)  ,
                        \nonumber \\
        g_{rr}(r) &=&  1  ,  \qquad  (0< r <d_m,  \quad l_c\leq r)  .
               	\label{eq:2052}
\end{eqnarray}	

 The  gradient   in Eq.(\ref{eq:2692}) is rewritten  as follows  (see Appendix.B) 
\begin{equation}
    H_{ef}=\int  g_{\mu\nu}\frac{\partial \widehat{f}^{\dagger} }{\partial x^{\mu}}  \frac{\partial \widehat{ f}}{\partial x^{\nu}}   d^3 x
	    + \int  W(x) \widehat{ f}^{\dagger} (x)  \widehat{ f}(x) d^3x   ,
               	\label{eq:20}
\end{equation}	
where a normalized  field  $\widehat{ f} (x ) \equiv |g(x)|^{1/4}f(x )$ has, in addition to the kinetic energy,  a  square of the  finite energy  as  follows  ($\mu, \nu = x,y,z$) \cite{orl} 
\begin{equation}
  W(x) = \frac{1}{4} \frac{\partial}{\partial x^{\mu}}
                 \left(g_{\mu\nu}\frac{\partial \ln |g|}{\partial  x^{\nu}} \right)
     +\frac{1}{16} g_{\mu\nu} \left(\frac{\partial \ln |g|}{\partial x^{\mu}} \right)
	                   \left(\frac{\partial \ln |g|}{\partial x^{\nu}} \right) ,
               	\label{eq:2051}
\end{equation}	
where $|g(x)|$ is a determinant of $g_{\mu\nu}(x)$. Using  Eq.(\ref{eq:2052}) and $g_{zz}=1$,   $W(r)$ of each quasi boson is given by
\begin{equation}
       W(r)= \frac{3}{4} \frac{1}{d_m^2} .
               	\label{eq:2055}
\end{equation}	
  Each quasi boson  should jump an  energy barrier $\epsilon _0 \equiv \sqrt{W(x)}$ when it is transversely excited.   The number of quasi bosons in  the transverse plane  within the coherence length   is $\pi (l_c/d_m)^2$. The sum of   $ \epsilon _0$   over quasi bosons  in this area  is given by
\begin{equation}
       \widehat{\epsilon} _0 = \frac{\sqrt{3}}{2} \pi  \left(\frac{l_c}{d_m}\right)^2 \frac{1}{d_m}  .
               	\label{eq:20571}
\end{equation}	
Owing to this  gap $\widehat{\epsilon} _0$,  the state described by $\widehat{ f}(x )$ remains in the  ground state even if it is  transversely  perturbed,  and  leads  to   $ \langle \widetilde{0} | \partial_{\mu}  \widehat{ f}(x) |\widetilde{0}\rangle =0$   within the coherent spatial  region.  Hence,  the state described by  $\varphi(x)$  also remains in  the ground state, leading   to  $ \langle \widetilde{0} |   \partial_{\mu} [ \bar{\varphi}(x) \gamma^{\mu} \varphi  (x)] |\widetilde{0}\rangle =0$ for the transverse direction.  This  exhibits a kind of rigidity of the physical vacuum, which is the origin of  the vacuum condensate $v_h$  in $ |(i\partial_{\mu}+gB_{\mu})(v_h+h_1+ih_2)|^2$  in Eq.(\ref{eq:01}).

\section{Massive gauge boson }
In the BEH model, the mass term  of gauge boson  $m_B^2B^{\mu}B_{\mu}= g^2 v_h^2B^{\mu}B_{\mu}$ is derived from the phenomenological coupling $ |(i\partial_{\mu}+gB_{\mu})(v_h+h_1+ih_2)|^2$.  In this paper, we derive this term microscopically.
  The physical vacuum  is not a simple system, and therefore     the response of  $ |\widetilde{0}\rangle$ to $B_{\mu}$  includes  a non-linear effect.  The minimal coupling of fermions  $L_0^{min}(x)=\bar{\varphi}(x)(i\partial_{\mu}+gB_{\mu})\gamma^{\mu}\varphi(x) $  changes its form  in  $|\widetilde{0}\rangle$  by  the  perturbation  of ${\cal H} _I(x)= g j^{\mu}(x) B_{\mu}(x)$    \cite{double}.   Consider a perturbation expansion of  $\int d^4xL_0^{min}(x)$  in powers of $g$ 
 \begin{eqnarray}
     \lefteqn{ \langle \widetilde{0}|  \int  d^4 x_1 L_0^{min}(x_1)
                                                    exp\left( i\int {\cal H} _I(x_2) d^4x_2 \right) |\widetilde{0}\rangle } 
                                                        \nonumber \\
     &&=   \langle \widetilde{0}|   \int d^4x_1 \bar{\varphi}(x_1) \gamma ^{\mu}[i\partial_{\mu} + gB_{\mu}(x_1)] \varphi(x_1)  |\widetilde{0}\rangle
                                                           \nonumber \\
            &&+     \langle \widetilde{0}|   \int d^4x_1  \bar{\varphi}(x_1) \gamma ^{\mu}[i\partial_{\mu} + gB_{\mu}(x_1)] \varphi(x_1)
                                                  ig \int d^4x_2 j^{\nu}(x_2) B_{\nu}(x_2)  |\widetilde{0}\rangle   
                                                                                   + \cdots      ,   
                                                                                               	 \label{eq:34}  
\end{eqnarray}

\subsection{ Gauge-boson's mass due to statistical gap}
(1)  In the last term of Eq.(\ref{eq:34}),  $B_{\nu}(x_2)$ is   correlated to $B_{\mu}(x_1)$,  yielding the following  two-point-correlation function between $ j^{\mu}(x_1)$  and $j^{\nu}(x_2)$
\begin{equation}
                        g^2 \int \langle \widetilde{0}|    \int  j_{\mu}(x_1)   d^2x_1  
                                                \int   j^{\nu}(x_2)    d^2x_2   |\widetilde{0}\rangle
                                                                  B^{\mu}(x_1)B_{\nu}(x_2) d^2x_1d^2x_2 
		                  						                	 \label{eq:35}
\end{equation}
  Since  these $x_1$ and $x_2$ are separated  microscopically,   a  distant observer  regards it as  a local phenomenon at   $X=(x_1+x_2)/2$. For such an observer, it is useful to  rewrite   $d^4x_1d^4x_2$ in Eq.(\ref{eq:35}) as $d^4Xd^4Y$.   \begin{equation}
        g^2  \int   \langle \widetilde{0}|   \int  j_{\mu}(Y) 
                                                  j^{\nu}(0) d^4Y  |\widetilde{0}\rangle \times  B^{\mu}(X)B_{\nu}(X) d^4X                                         
                                               	 						                	 \label{eq:31.8}                                              	
\end{equation}
The relative motion along   $Y=x_2-x_1$  is indirectly observed in the coefficient of $B^{\mu}(X)B_{\nu}(X)$
 
(2)  Because the gauge field  is a transverse one,  the excitation of fermions  induced by this $B_{\mu}(x)$  is a transverse one as well.  The correlation of currents in Eq.(\ref{eq:31.8}) implies a correlation of the   center-of-mass motions of different massless fermion-antifermion  pairs. Because they are  strongly influenced by   Bose statistics,  quasi bosons  move   parallel to each other coherently  in the response to $B_{\mu}$,  and  correlation of different  currents is nonzero only  in  the same direction  as follows
\begin{eqnarray}
	  \langle \widetilde{0}|   j_{\mu}(Y)  j^{\nu}(0)   |\widetilde{0}\rangle & \Rightarrow &  \langle \widetilde{0}|   j_{\mu}(Y)  j^{\mu}(0)  |\widetilde{0}\rangle \quad  for \quad \mu=\nu .	                                                           		   	 \nonumber \\
          \langle \widetilde{0}|   j_{\mu}(Y)  j^{\nu}(0)   |\widetilde{0}\rangle  &\Rightarrow  & 0  \qquad  \qquad \qquad \qquad  for \quad  \mu \ne \nu .	
	                                                           		   	 \label{eq:3599}
\end{eqnarray}   
 \begin{figure}
	 	\centering
\includegraphics [bb=0 0 657 407, scale=0.44] {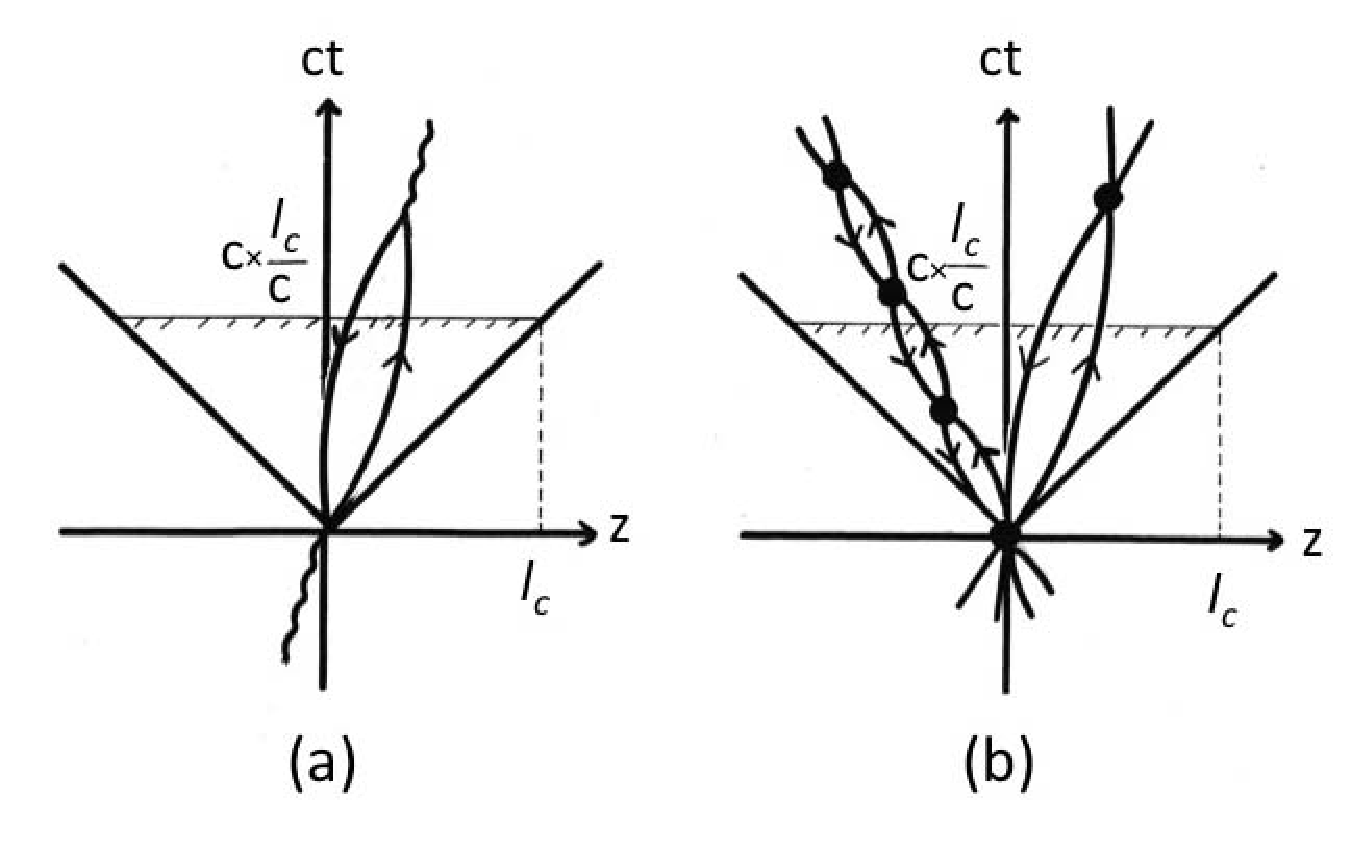}
\caption{ (a)  In the non-Fock world, the gauge boson in $B_{\mu}$ (a wavy line)  induces vacuum polarization of massless fermion and antifermion  only outside the coherent space-time region specified by $l_c$  (a shaded area). (b)  In the real world, the chain of creation and annihilation of massive  fermion and antifermion   constitutes  the Higgs-like excitation $H(x)$, which is  induced  both inside and  outside  this  region .  }
\label{fig.3}
\end{figure}
Within the coherence length  $l_c$, $\varphi  (x)$ in the current  remains in the ground state, and  $ \langle \widetilde{0} |   \partial_{\mu}[ \bar{\varphi}(Y)\gamma^{\mu}   \varphi  (Y)] |\widetilde{0}\rangle =0$  holds for the transverse component of  $Y$.   The correlation of currents in Eq.(\ref{eq:31.8})  is reduced to
\begin{eqnarray}
		 \langle \widetilde{0} |  j_{\mu}(Y)  j^{\mu}(0)  |\widetilde{0}\rangle
               &=&  \langle \widetilde{0} | \left( j_{\mu}(0)+  \left[\frac{\partial}{\partial Y_{\mu}} [\bar{\varphi}(Y) \gamma_{\mu}
                        \varphi  (Y)]\right]_{Y_{\mu}=0}  Y^{\mu}+ \cdots \right)     j^{\mu}(0)  |\widetilde{0}\rangle
                                               \nonumber \\
                & \Rightarrow &  \langle \widetilde{0} |  j_{\mu}(0) j^{\mu}(0)   |\widetilde{0}\rangle.
		 \label{eq:111}
\end{eqnarray}
Using  $j^{\mu}(0)=( \varphi^{\dagger} \varphi, i \varphi^{\dagger} \gamma^0 \mbox{\boldmath $\gamma$} \varphi )$, we obtain 
$ \langle \widetilde{0} | j_{0}(0) j^{0}(0) |\widetilde{0}\rangle =  \langle \widetilde{0} | j_{1}(0) j^{1}(0) |\widetilde{0}\rangle =  \langle \widetilde{0} | j_{2}(0) j^{2}(0) |\widetilde{0}\rangle =
 \langle \widetilde{0} | j_{3}(0) j^{3}(0) |\widetilde{0}\rangle = \langle \widetilde{0} | [ \varphi ^{\dagger} (0) \varphi  (0)]^2  |\widetilde{0}\rangle $.
 Hence,  Eq.(\ref{eq:31.8}) is interpreted as the  mass term of gauge boson even after the Fock vacuum is restored 
\begin{equation}
	 \frac{1}{2}m_B^2  \int  B^{\mu}(X) B_{\mu}(X) d^4X  ,
	                                                           		   	 \label{eq:36}
\end{equation}
where  $m_B^2$   is given by
  \begin{equation}
	m_B^2=  2g^2  \langle \widetilde{0}|  \int_{Y\in  Z_c} [ \varphi ^{\dagger} (0) \varphi  (0)]^2 d^4Y |\widetilde{0}\rangle 	 .	
	                                                           		   	 \label{eq:3601}
\end{equation}
The {\bf coherent space-time region} $Z_c$  in Eq.(\ref{eq:3601}) is the inside of a small  light-cone   as illustrated  in  Figure \ref{fig.3} for one spatial direction. Its  spatial  length is less than $l_c$, and time width is  smaller  than $l_c/c$, so that  the causal relation is  possible between two spatial  ends separated by $l_c$.  The 4-dimensional volume of  coherent space-time region $ \int_{Y\in Z_c}d^4Y=l_c^2\times\frac{1}{2} [l_c \times c(l_c/c)]\times 2=l_c^4$  in  Figure \ref{fig.3}  is Lorentz invariant.   Since transverse  excitations are suppressed  in $Z_c$, the vacuum polarization induced by $B_{\mu}$ occurs only   outside  this region as shown in Figure \ref{fig.3}(a).

 The $\varphi(x)$ in Eq.(\ref{eq:3601})  has a normalization volume $V=d_m^3$ as  in  Eq.(\ref{eq:211}).   Every fermion and antifermion in $ |\widetilde{0}\rangle $ make the same contribution to  $ \langle \widetilde{0}| [ \varphi ^{\dagger} (0) \varphi  (0)]^2 |\widetilde{0}\rangle $ as follows 
 \begin{eqnarray}
   	&\,&  \langle  \widetilde{0}|  \frac{1}{d_m^6}  \sum_{p,s}
              \left( [a^{s\dagger} (\mbox{\boldmath $p$})u^{s\dagger} (p)+ b^{s }(-\mbox{\boldmath $p$})v^{s\dagger} (-p)] 
                  [a^s(\mbox{\boldmath $p$})u^s(p)+ b^{s\dagger }(-\mbox{\boldmath $p$})v^s(-p)] \right)^2 |\widetilde{0}\rangle  
                                  \nonumber \\
         &=&
   	 \frac{1}{d_m^6}  \langle \widetilde{0}|  \sum_{p,s} [b^{s }(-\mbox{\boldmath $p$}) b^{s\dagger }(-\mbox{\boldmath $p$})]^2
	                                                           + \sum_{p,s} [a^{s\dagger} (\mbox{\boldmath $p$}) a^s(\mbox{\boldmath $p$})]^2
	                           \nonumber \\
	 &+& \sum_{p,s} b^{s }(-\mbox{\boldmath $p$}) a^s(\mbox{\boldmath $p$}) a^{s\dagger} (\mbox{\boldmath $p$}) b^{s\dagger }(-\mbox{\boldmath $p$})
	      + \sum_{p,s} a^{s\dagger} (\mbox{\boldmath $p$}) b^{s\dagger }(-\mbox{\boldmath $p$}) b^{s }(-\mbox{\boldmath $p$}) a^s(\mbox{\boldmath $p$}) 
	                                            \,   |\widetilde{0}\rangle  
                                  \nonumber \\
         &=& 2 \times  \frac{1}{d_m^6} \prod_{p,s}  ( \cos^2\theta _{\mbox{\boldmath $p$}} + \sin ^2\theta _{\mbox{\boldmath $p$}} ) = \frac{2}{d_m^6}.
	                                                           		   	 \label{eq:360002}
\end{eqnarray}
where $u^{s\dagger} (p) u^s(p)= v^{s\dagger} (-p)v^s(-p) =1 $.  The gauge boson's mass  for  $l_c^4$ of  $Z_c$ in 
Eq.(\ref{eq:3601}) is given  by
 \begin{equation}
   m_B^2=  g^2 \frac{4}{d_m^6}  \int_{Y\in  Z_c}  d^4Y=    g^2 \left( \frac{2 l_c^2 }{d_m^3} \right )^2 .
	                                                           		   	 \label{eq:361}
\end{equation}
This $2 l_c^2/d_m^3$ is the origin  of  vacuum condensate  $v_h$ in $m_B^2=g^2v_h^2$ of  the BEH model, which is about  the same as $\widehat{\epsilon} _0$ in Eq.(\ref{eq:20571}).

\subsection{Goldstone mode}

(1)  The Goldstone mode exists  in the last term of  Eq.(\ref{eq:34}).  For the first-order term of $B_{\nu}(x_2)$ coupled to  $\bar{\varphi}(x_1) \gamma _{\mu} i\partial^{\mu} \varphi(x_1)$,  integrate it  over $x_1$  by parts. Since  $ \varphi(x_1)$ vanishes at $x_1 \rightarrow \infty$,  we obtain  two types of terms, one including $i\partial^{\mu}\bar{\varphi}(x_1)\gamma _{\mu} \varphi(x_1)$,  and  the other including $\partial^{\mu} |\widetilde{0}\rangle$. The latter is given by
\begin{eqnarray}
	&\,& g \langle \widetilde{0}|    \int d^4x_1    j_{\mu}(x_1)   \int d^4x_2   j^{\nu}(x_2) B_{\nu}(x_2)   \partial^{\mu} |\widetilde{0}\rangle 
	                         \nonumber \\
         &+& g  \partial^{\mu} \langle \widetilde{0}|    \int d^4x_1       j_{\mu}(x_1)    \int d^4x_2   j^{\nu}(x_2) B_{\nu}(x_2)    |\widetilde{0}\rangle  .
	                                    						                	 \label{eq:3741}
\end{eqnarray}
Because the  vacuum $ |\widetilde{0}\rangle$  in Eq.(\ref{eq:16})  has an explicit $x$-dependence in the phase  $\alpha(x)$,   $\partial^{\mu} |\widetilde{0}\rangle $  contains 
 $ \partial^{\mu}\alpha (x)$.  The distant observer regards Eq.(\ref{eq:3741}) as representing  a local phenomenon at $X$, and   rewrites it  using $d^4x_1d^4x_2=d^4Xd^4Y$. 
  After extracting $m_B^2$ from Eq.(\ref{eq:3741}) as in   Eq.(\ref{eq:31.8}), we obtain 
\begin{equation}
	\frac{i }{2g} m_B^2 \int B_{\mu}(X)\partial ^{\mu} \alpha (X) d^4X 
	                \equiv  \frac{m_B}{\sqrt{2}} \int B_{\mu}(X) \partial^{\mu} G(X) d^4X   .
	                              						                	 \label{eq:38}
\end{equation}
Here the Goldstone mode is defined as $G (X)\equiv   i (\sqrt{2}g)^{-1}m_B \alpha (X)$. This definition takes over the bound-state realization of the Goldstone mode from  \cite{nam}.
(While  the BEH model derives the coupling of the Goldstone mode $h_2$  to $B_{\mu}$   from 
 $ |(i\partial_{\mu}+gB_{\mu})(v_h+h_1+ih_2)|^2$,  the above coupling  comes from  the  response of the vacuum $ |\widetilde{0}\rangle$ to $B_{\mu}$.)

(2) In the system without the long-range force, the global phase-rotation  of $\alpha (X)$  requires no energy,  and therefore the propagator of the Goldstone mode is given by
\begin{equation}
	 \int \frac{d^4X}{(2\pi)^4} \langle \widetilde{0}| T[ G(X) G (0) ]  |\widetilde{0}\rangle e^{iqX} = \frac {i} { q^2 }  .
		 \label{eq:381}
\end{equation}
 However, the long-range force mediated by the gauge boson  prevents  the  free  rotation of the global phase, then preventing  the Goldstone mode. This discrepancy is resolved  by the generation of the  gauge-boson's mass that converts  the long-range force into a short-range one.  The Fourier transform of Eqs.(\ref{eq:36}) and (\ref{eq:38})   are given by $(m_B^2/2)B^{\mu}(q)B_{\mu}(q)$  and $(m_B/\sqrt{2})q^{\mu} G(q) B_{\mu}(q)$, respectively. Following the usual way, regard the latter as a  perturbation to the former,  and the second-order perturbation is obtained as
\begin{equation}
	\frac{1}{2} B^{\mu}(q) \left [i m_B^2g^{\mu\nu} - m_Bq^{\mu}  \frac{i}{q^2} m_Bq^{\nu} \right] B_{\nu}(q)	
	                     = \frac{i}{2} m_B^2 \left( g^{\mu\nu} - \frac{q^{\mu}q^{\nu}}{q^2} \right) B^{\mu}(q)B^{\nu}(q).
		 \label{eq:382}
\end{equation} 
Adding Eq.(\ref{eq:382})  to the Fourier transform of $-\frac{1}{4} F^{\mu\nu}F_{\mu\nu}$,  we  obtain  an  inverse  of  the resulting matrix, 
\begin{equation}
		       D^{\mu\nu}(q) = \frac{-i}{q^2-m_B^2} \left( g^{\mu\nu} - \frac{q^{\mu}q^{\nu}}{q^2} \right) 
		                         		       \equiv i D(q^2) \left( g^{\mu\nu} - \frac{q^{\mu}q^{\nu}}{q^2} \right),
		 \label{eq:384}
\end{equation}
which is the propagator of the massive gauge boson in the Landau  gauge.  The response of the physical vacuum gives  additional terms to $L_0(x)$  as  $(m_B^2/2)B^{\mu}(q)B_{\mu}(q)$, $(m_B/\sqrt{2}) B_{\mu}(x)  \partial^{\mu} G (x)$ and 
$( \partial_{\mu}G(x))^2$.    

(3)  In the BEH model, the  coupling of the Goldstone mode to fermions   comes explicitly  from the Yukawa interaction. In the present model,   such a  coupling appears  implicitly in the first term of the right-hand side in  Eq.(\ref{eq:34}).   For the zeroth-order term of $B_{\mu}$, the  integration  over $x_1$ by parts  yields  two types of  terms, one including  $i\partial^{\mu}\bar{\varphi}(x_1)\gamma _{\mu} \varphi(x_1)$,  and the other including  $\partial^{\mu} |\widetilde{0}\rangle$. The latter term is given by 
\begin{equation}
		i \langle \widetilde{0}|   \int d^4x_1 j^{\mu}(x_1) \partial_{\mu} |\widetilde{0}\rangle  + i  \partial_{\mu} \langle \widetilde{0}|   \int d^4x_1 j^{\mu}(x_1) |\widetilde{0}\rangle    .
		 \label{eq:385}
\end{equation}
Since  $ \partial_{\mu} |\widetilde{0}\rangle =  \partial_{\mu}\alpha(x)  |\widetilde{0}\rangle $ contains the Goldstone mode $G(x)= i (\sqrt{2}g)^{-1}m_B\alpha(x)$,  Eq.(\ref{eq:385}) gives  the coupling of the  Goldstone mode $G$  to fermion
\begin{equation}
		 \frac{\sqrt{2}g}{m_B}  \langle \widetilde{0}|   \int d^4x_1  
		            \bar{\varphi} (x_1)\gamma^{\mu} \varphi (x_1)  \partial_{\mu}G(x_1) |\widetilde{0}\rangle .
		 \label{eq:3851}
\end{equation}
Here,  a new term  $(\sqrt{2}g/m_B) \bar{\varphi} (x)\gamma^{\mu} \varphi (x) \partial_{\mu}G(x)$  is added to  $L_0(x)$, which is different from  $g(m_f/m_B) \bar{\varphi} (x) \varphi (x)h_2(x)$ led by Yukawa coupling  in the BEH model.

\section{Higgs boson  }
In the BEH model,  the Higgs boson  appears in a strange manner, that is,  $-\mu ^2|h|^2$ in $L_h(x)$ of  Eq.(\ref{eq:01}) plays double roles. The first is the  generation of the broken-symmetry vacuum by switching  the sign of $\mu^2$. The second is giving the  Higgs boson  a mass $m_H$. However, the  former is concerned with the global property of the world, and the latter is concerned with one-particle property. It seems strange that such different scale of things are described by the same  parameter. In the present model,  since the vacuum   $ |\widetilde{0}\rangle$ has  the kinematical origin,  we  have an option to  assume  a possible  interaction for deriving $m_H$  without caring about symmetry breaking.

One of the important predictions of  the BEH model  for  the electroweak interaction  is that the strength of the Higgs's coupling to fermions is proportional to the fermion's mass $m_f$ as in  the Yukawa interaction $ (m_f/v_h)(v_h+h_1+ih_2) \bar{\varphi}\varphi $,   which is confirmed  by experiments.  Hence,  the next step is  to go   beyond the  mean-field $U_0$  in  Eq.(\ref{eq:751}).   In the real Fock world,  the massive gauge boson is transmitted from one  massive   fermions to the other. Fermions and antifermions   exist not  as a pair but as an  independent  particle, and their creation or annihilation   is defined by a field \cite{massive}
  \begin{equation}
		i \psi(x)= \int \frac{d^3p}{(2\pi)^3} \sum_{s} \frac{1}{\sqrt{2E_p}}
		                     \left[ \widehat{a}^s(\mbox{\boldmath $p$})\widehat{u}^s(p)e^{-ipx}
                                                  + \widehat{b}^{s\dagger }(\mbox{\boldmath $p$})\widehat{v}^s(p)e^{ipx} \right]    .
\end{equation}
When   massive fermions in scattering do not have  enough  kinetic energy to generate the on-shell  massive gauge boson,   the non-relativistic limit  is a good approximation  in observer's reference frame.   The first approximation of the massive  fermion-fermion scattering amplitude  \cite{gol}  is given by 
  \begin{equation}
		i {\cal M }= (-ig)^2 \bar{\widehat{u}}^{s'}(p_1')  \gamma^{\mu} \widehat{u}^s(p_1) 
		                    \frac{-ig^{\mu \nu}}{q^2-m_B^2}  \bar{\widehat{u}}^{s'}(p_2')  \gamma_{\nu} \widehat{u}^s(p_2)   .
		                                    \label{eq:1711}
\end{equation}
Taking only $\mu=\nu=0$,  and using $q^2=(p_1'-p_1)^2 \ll m_B^2$ and  $ \widehat{u}^{s'}(p_1')\gamma^0 \widehat{u}^s(p_1) \simeq  2m_f \delta^{ss'}$ in Eq.(\ref{eq:1711}), we obtain ${\cal M}=4g^2(m_f^2/m_B^2)$. (For fermion-antifermion, and antifermion-antifermion scattering, the same ${\cal M }$ is obtained.) In collision experiments,  the effective  interaction that  leads to this amplitude  induces a  {\it collective motion  of the massive fermion and antifermion}. In place of   Eq.(\ref{eq:751}), we obtain
\begin{equation}
            \bar{\psi}(x) \left[  i \sla{\partial }  + \widehat{g} H(x)  \right]  \psi(x) ,   
			                                    \label{eq:1713}
\end{equation}
where  the scalar field $H(x)$  denotes this collective motion, and  $\widehat{g}$ is  given by 
\begin{equation}
            \widehat{g} =  \frac{m_f}{m_B}  g .  
			                                    \label{eq:1714}
\end{equation}
In  $L'_h(x)$ of Eq.(\ref{eq:011}),  the coupling constant $m_f/v_h$  of the  Higgs boson  to fermions is,   using the relation $m_B=gv_h$,  equal to $(m_f/m_B)g$,  which  agrees with  this $\widehat{g}$. Hence, there arises  a possibility that the scalar field $H(x)$ in Eq.(\ref{eq:1713}) represents  the Higgs field.  Here we consider the possibility that  the Higgs boson  is not an elementary particle, but a quasiparticle representing a collective excitation. 

The simplest collective excitation   is  a series  of  creations and annihilation  of massive  fermion and antifermion that   propagate  in space as  in  Figure \ref{fig.3}(b), where   black circles  represent  a vertex made of Eq.(\ref{eq:1713}).   The propagator of the Higgs  field  is given by 
\begin{equation}
 \int \frac{d^4x}{(2\pi)^4}   \langle \widetilde{0}| T[ H(x) H(0) ] | \widetilde{0}\rangle  e^{iqx} =\frac{1}{q^2\left[1- \chi(q^2)\right]}      .
						                	 \label{eq:48}
\end{equation} 
 Since  this  excitation  is not a transverse one,  there is no energy gap in its  excitation spectrum. Here we define a finite {\bf upper end} $\xi _c$ of energy-momentum of the excitation  as in the non-relativistic physics. 
 The self energy of the Higgs  boson  is  given by 
\begin{equation}
i q^2\chi (q^2)= (-i  \widehat{g})^2 (-1) \int ^{\xi_c}_0 \frac{d^4p}{(2\pi)^4} tr\left[ \frac{i}{\sla{p}-m_f} \frac{i}{\sla{p}+\sla{q}-m_f} \right]    ,
						                	 \label{eq:461}
\end{equation}
in which $\gamma ^{\mu}$ matrix is not there. This integral is not   a loop of the relativistic  process extending to infinite momentum,  but an  integral over $p$  from zero to this  $\xi_c$.  According to  the ordinary rule, we  use a new variable $l=p+xq$.  The upper end in the integral over $l$ is  $\sqrt{p^2+2xp\cdot q +x^2q^2}$,  which depends on the relative direction of $p$ to $q$. Since the sign of $p\cdot q$ oscillates  between positive and negative, we use a mean value  $\sqrt{p^2+x^2q^2}$ for simplicity.   Hence,  using an Euclidian 4-momentum $l_E$ as  $l^2= -l_E^2$,  we obtain   
\begin{equation}
	q^2\chi (q^2)= -4   \widehat{g}^2 \int^1_0 dx \int \frac{d\Omega_4}{(2\pi)^4}\int_{\sqrt{x^2q^2}}^{\sqrt{\xi_c^2+x^2q^2}} l_E^3 dl_E
                           \left[ \frac{-l_E^2} {(l_E^2+\Delta )^2}  + \frac{\Delta}{ (l_E^2+\Delta )^2}  \right]    ,
						                	 \label{eq:46}
\end{equation}
where $\Delta=m_f^2-x(1-x)q^2$.   If we define the  following  integral
\begin{equation}
	  I(m,n)\equiv \int l_E^m(l_E^2 + \Delta)^n dl_E    ,
						                	 \label{eq:465}
\end{equation}
the indefinite integrals over $l_E$ in Eq.(\ref{eq:46}) are  decomposed as follows
\begin{equation}
	I(5,-2) - \Delta \times I(3,-2) =  I(1,0) + 2\Delta ^2 \times I(1,-2) - 3\Delta \times  I(1,-1)   ,
						                	 \label{eq:466}
\end{equation}
where
\begin{equation}
	I(1,0) = \frac{1}{2} l_E^2,  \quad  I(1,-2)=-\frac{1}{2(l_E^2+\Delta)},   \quad I(1,-1) = \frac{1}{2} \ln |l_E^2 +\Delta| .
						                	 \label{eq:467}
\end{equation}
The definite  integral  over $l_E$ in Eq.(\ref{eq:46}) yields  
\begin{eqnarray}
q^2 \chi(q^2)= \frac{ \widehat{g}^2}{4\pi^2}  \xi_c ^2
    &-&\frac{ \widehat{g}^2}{2\pi^2}   \int_{0}^{1}dx  \Delta ^2  \left( \frac{1}{ \xi_c ^2+ x^2q^2+\Delta}  - \frac{1}  {x^2q^2+\Delta}\right)
                          \nonumber \\
           &-& \frac{ \widehat{g}^2}{2\pi^2}   \int_{0}^{1} dx \frac{3}{2}  \Delta \ln \left| 1+ \frac{ \xi_c ^2}{x^2q^2+\Delta} \right|  .
	                              						                	 \label{eq:471}
\end{eqnarray}
 With this $\chi(q^2)$,    the mass $m_H$ of the  Higgs boson  is defined  as $\chi(q^2) \simeq m_H^2/q^2$ at $q^2 \rightarrow 0$.    Since $\Delta \rightarrow m_f^2$ at $q^2 \rightarrow 0$,   the integrals over $x$ in the second and third terms of the  right-hand side of  Eq.(\ref{eq:471}) have following limits at $q^2 \rightarrow 0$ 
\begin{equation}
    \int_{0}^{1}dx   \frac{1}{ \xi_c ^2+ x^2q^2+\Delta}  \rightarrow \frac{3}{4(\xi_c ^2+m_f^2)} ,
						                	 \label{eq:474}
\end{equation}
\begin{equation}
   \int_{0}^{1} dx   \ln \left| 1+ \frac{ \xi_c ^2}{x^2q^2+\Delta} \right| \rightarrow \ln \left| \frac{\xi_c ^2+ m_f^2}{m_f^2}\right|.
						                	 \label{eq:476}
\end{equation} 
Plugging   Eqs.(\ref{eq:474}) and (\ref{eq:476})  into Eq.(\ref{eq:471}), and using it in   Eq.(\ref{eq:48}),  the mass $m_H$  is given by
\begin{equation}
  m_H^2 =  \frac{ \widehat{g}^2}{4\pi^2} \left[ \xi_c^2+  \frac{3}{2} m_f^2 \left(1 -  \frac{m_f^2}{\xi_c^2+m_f^2}\right) 
                  -  3 m_f^2\ln \left ( \frac{ \xi_c ^2+m_f^2}{m_f^2}\right ) \right]  .
						                	 \label{eq:4854}
\end{equation}
The Higgs mass  is related to  $\xi_c$, $m_f$,  and $m_B$ in $\widehat{g}$. This $\xi_c$ is  to be  determined by experimental value of $m_H$. Once a finite  value of $\xi _c$  is obtained, divergence that we face   in further  renormalization  is not a  quadratic, but a  logarithmic divergence, because we need not  use  the Higgs potential.

The Higgs field  is described by the following effective  Lagrangian density
\begin{equation}
          (\partial_{\mu} H)^2 -  m_H^2 H^2  +  \frac{m_f}{m_B} g \bar{\psi}\psi H .
						                	 \label{eq:49}
\end{equation}
Whereas  the masses  of  fermion and gauge boson have  their  origin in the non-Fock vacuum,   the Higgs boson  is   a many-body phenomenon in the Fock vacuum. 

Compared to   the BEH model  $L_0(x)+L'_h(x)$,  this picture of Higgs boson  has  following   features. 

(1)  In the experimental tests of Higgs particle,  the Higgs coupling  to massive gauge bosons is  an important channel.    In the BEH model,  the gauge coupling $ |(i\partial_{\mu}+gB_{\mu})(v_h+h_1+ih_2)|^2$  predicts a  direct coupling of  the  massive gauge boson  to the  Higgs field  $h_1$ or Goldstone mode $h_2$  as follows
\begin{equation}
     g^2v_h^2 B^{\mu}B_{\mu} \left(1+\frac{h_1}{v_h} \right) ^2 + g^2 B^{\mu}B_{\mu} h_2^2 + 2gB^{\mu} (h_1\partial _{\mu}h_2+ h_2\partial _{\mu}h_1 ) +(c.c)  .
						                	 \label{eq:502}
\end{equation}
In the present model,   the Higgs field  $H$ is not an elementary field but represents collective excitation, so that  gauge coupling  of $H$ with the same  $g$  does not exist.  Instead of $m_B^2B^{\mu}B_{\mu} (1+h_1/v_h)^2 $,  the effective  coupling of $H$ to $B_{\mu}$  arises in  the perturbation process induced by  $\widehat{g} \bar{\psi}\psi H$ and $g \bar{ \psi} \gamma^{\mu}\psi B_{\mu}$,  in which such  couplings are   relayed by  $\bar{\psi}$ and $\psi$. The  gauge coupling  of $h_1$ and $h_2$ in the BEH mechanism may be a useful phenomenology if  $g$ is adjusted properly.

(2)  In the GWS model,  Higgs and Goldstone  couplings to fermions are  proportional to fermion's mass.  In the present model, 
such couplings are  made  by $ (m_f/ m_B) g \bar{\psi}\psi H $  in Eq.(\ref{eq:49}) and $(\sqrt{2}g/m_B) \bar{\psi}  \gamma^{\mu} \psi   \partial_{\mu} G$   in Eq.(\ref{eq:3851}),  respectively.     For   the Goldstone's   coupling  to the fermion, it    arises  from the  vacuum $|\widetilde{0}\rangle$ itself.  The  Yukawa interaction $ (m_f/v_h)(v_h+h_1+ih_2) \bar{\varphi}\varphi $  is  a simple phenomenology for  the response of  physical vacuum to the gauge field.

\section{Discussion}

 The  following   total   Lagrangian  density  $\widetilde{L}(x)$ on the Fock space is obtained.  After rewriting $\varphi$ and $\bar{\varphi}$  with $\psi$ and $\bar{\psi}$,  $\widetilde{L}(x)$ is given by
\begin{eqnarray}
    \widetilde{L}(x) = &-& \frac{1}{4}F^{\mu\nu}F_{\mu\nu}    
                          +  \bar{ \psi}(i\partial_{\mu}+gB_{\mu})\gamma^{\mu}\psi - m_f\bar{\psi}\psi 
                                                        \nonumber \\
              &+& \frac{1}{2} m_B^2 B^{\mu} B_{\mu}  + \frac{m_B}{\sqrt{2}} B_{\mu} \partial^{\mu} G + ( \partial_{\mu} G)^2
                              +  \frac{\sqrt{2}g}{m_B}    \bar{\psi}  \gamma^{\mu} \psi\partial_{\mu} G
                                  \nonumber \\
               &+& (\partial_{\mu} H)^2 - m_H^2  H^2  +  \frac{m_f}{m_B} g \bar{\psi}\psi H  .
		 \label{eq:50}
\end{eqnarray}
 In this $ \widetilde{L}(x)$, the double role of the Higgs potential  is dissolved, and each  role  is played  by  each physical process.  Fermion's mass  and the vacuum condensate are not related.  Chiral-symmetry  breaking is not induced by the Higgs potential, but occurs in the restoration  process of the Fock vacuum. The  vacuum condensate $v_h$  arises from  the statistical gap in the transverse  excitation.   Whereas  the Higgs and Goldstone fields in the BEH model are defined  by the Higgs potential,  the Higgs boson   represents  the many-body phenomenon in the real world, and the Goldstone field arises  from the phase factor  in the vacuum $|\widetilde{0}\rangle$.   Parameters $\mu$ and $\lambda$ in the Higgs potential are replaced by parameters that have   physical  interpretations, 
\begin{eqnarray}
    m_B^2 &=& g^2 \left( \frac{2 l_c^2}{d_m^3}\right)^2 , \qquad (=g^2 \frac{\mu^2}{2\lambda} )
                                                            \nonumber \\
               m_f &=& |U_0|.
                   \nonumber \\                                          
              m_H^2 &=& \frac{ \widehat{g}^2}{4\pi^2} \left[ \xi_c^2+  \frac{3}{2} U_0^2 \left(1 -  \frac{U_0^2}  {\xi_c^2+U_0^2}\right) -  3 U_0^2\ln \left ( 1+ \frac{ \xi_c ^2}{U_0^2}\right ) \right] ,  \qquad (= 2\mu^2)
                                  \label{eq:501}
\end{eqnarray}
where $ \widehat{g}= (m_f/m_B)g = |U_0| (d_m^3/\sqrt{2} l_c^2)$.

The present model  has implications  for  some fundamental problems that lie  in the BEH model.

(a)   The BEH mechanism, before  applied to the electroweak interaction as a part of the GWS model,  was nothing but a theoretical possibility. When  applied to the electroweak  interaction,  this mechanism   became a realistic model, and 
   the  mass acquisition  not only for  gauge boson but also for fermion became necessary.  The Yukawa interaction was  a useful phenomenological resolution to it, but the origin  of the vacuum condensate $v_h$ remains to be proved.  A  concrete image of $v_h$ can be obtained  when the distribution  of  massless fermion and antifermion  in the non-Fock vacuum   is considered  in position space.

(b) Since  the present model  does not  assume  the Higgs potential $-4\lambda v_h h_1^3 -\lambda h_1^4$, the quadratic divergence does not occur in the perturbation calculation,  provided that  $\xi_c$ is  determined using the experimental value of $m_H$ in Eq.(\ref{eq:501}).  The divergence we must renormalize  is only logarithmic one.   The present model  proposes a solution to  the fine-tuning problem of the quadratic divergence.

(c)  According to the  lattice model,   which  strictly preserves  local gauge invariance  at each stage of argument,  the vacuum expectation value (VEV)  of the gauge-dependent quantity vanishes when  calculated   without gauge fixing. Hence,   if $h(x)$ follows   $h(x) \rightarrow h(x)\exp(i\theta (x))$ under $A_{\mu}(x) \rightarrow  A_{\mu}(x) -ie^{-1}\partial_{\mu}\theta(x)$,  $\langle h(x) \rangle=0 $ is unavoidable.  (Elitzur-De Angelis-De Falco-Guerra theorem) \cite{eli}\cite{gue}.  This is because the local character of gauge symmetry effectively breaks the connection  in the   degrees of freedom defined  at  different space-time points.   If the  Higgs boson  is an elementary  particle,  the finite  $\langle h(x) \rangle $ that  depends on the gauge-fixing procedure   does not match its fundamental nature.  Instead of  $v_h=\langle h(x) \rangle$, the  vacuum is characterized by $ \langle \widetilde{0}|  \int [ \varphi ^{\dagger} (0) \varphi  (0)]^2 d^4Y  |\widetilde{0}\rangle$ in Eq.(\ref{eq:3601}).   Because   this condensate   is  gauge invariant,  there is no need to worry about  the  vanishing of its VEV.

(d)    In the Higgs potential,  $\lambda |v_h+h_1+ih_2|^4$ predicts the triple and quartic self-couplings of the  Higgs boson.   More complex many-body effects  than that  in Figure \ref{fig.3}(b)  may correspond to such self couplings.  So far, the agreement of the GWS model to experiments  is satisfactory.  When  more precise measurements are performed,  however,  there is a possibility of deviation, especially for  light quarks and leptons of the first and second generations. 
The above  $\widetilde{L}(x)$ predicts some  different results   from those by  the BEH model.  The next  subject is to  extend it  to the electroweak interaction.

\begin{appendices}

\section{The vacuum $|\widetilde{0}\rangle $  satisfying  $\widetilde{a}^s(\mbox{\boldmath $p$})|\widetilde{0}\rangle =\widetilde{b}^s(-\mbox{\boldmath $p$})|\widetilde{0}\rangle =0$} \label{secA1}

 The vacuum satisfying   $\widetilde{a}^s(\mbox{\boldmath $p$})|\widetilde{0}\rangle =\widetilde{b}^s(-\mbox{\boldmath $p$})|\widetilde{0}\rangle =0$    is  as follows. 
 In the  expansion  
\begin{equation}
                    e^{-iK} F e^{iK} = F + [-iK,F]+  \frac{1}{2!} \left[-iK, [-iK,F] \right] + \cdots ,
						                	 \label{eq:10}
\end{equation}
we regard an operator  $a^s(\mbox{\boldmath $p$})$ or $b^s(-\mbox{\boldmath $p$})$ as $F$,  and the  following operator as $K$,
\begin{equation}
     i \sum_{p,s}  \theta_{\mbox{\boldmath $p$}} [b^{s\dagger}(-\mbox{\boldmath $p$})a^{s\dagger}(\mbox{\boldmath $p$})-a^s(\mbox{\boldmath $p$})b^s(-\mbox{\boldmath $p$})] .
						                	 \label{eq:951}
\end{equation} 
Hence,   Eqs.(\ref{eq:8}) and (\ref{eq:9}) can be rewritten  in a compact form
\begin{equation}
		\widetilde{a}^s(\mbox{\boldmath $p$})= e^{-iK}a^s(\mbox{\boldmath $p$})e^{iK}    , \quad					            
		\widetilde{b}^s(-\mbox{\boldmath $p$})= e^{-iK}b^s(-\mbox{\boldmath $p$})e^{iK}   .
						                	 \label{eq:12}
\end{equation}
The vacuum $|\widetilde{0}\rangle$  satisfying  $\widetilde{a}^s(\mbox{\boldmath $p$})|\widetilde{0}\rangle =\widetilde{b}^s(-\mbox{\boldmath $p$})|\widetilde{0}\rangle =0$  is simply expressed as  $|\widetilde{0}\rangle = e^{-iK}|0\rangle$.   Hence, we obtain
\begin{eqnarray}
		|\widetilde{0}\rangle&=& 
	\exp\left( \sum_{p,s}\theta_{\mbox{\boldmath $p$}} [b^{s\dagger}(-\mbox{\boldmath $p$})a^{s\dagger}(\mbox{\boldmath $p$})
		                                        -a^s(\mbox{\boldmath $p$})b^s(-\mbox{\boldmath $p$}) ]\right ) |0\rangle  .
		                                         \nonumber \\
         	&=& \prod_{p,s} \left[ \sum_{n} \frac{1}{n!}  \theta_{\mbox{\boldmath $p$}}^n   [b^{s\dagger}(-\mbox{\boldmath $p$})a^{s\dagger}(\mbox{\boldmath $p$})
		                                        -a^s(\mbox{\boldmath $p$})b^s(-\mbox{\boldmath $p$})]^n \right]    |0\rangle.        
		                                           		                                                 \label{eq:13}
\end{eqnarray}
Each fermion and antifermion    obey Fermi statistics,  and therefore only a single particle can occupy each state. The sum over $n$  in Eq.(\ref{eq:13}) is written  for each $\mbox{\boldmath $p$}$ as follows
\begin{eqnarray}
	\sum_{n} \frac{\theta^n}{n!} (b^{\dagger}a^{\dagger}-ab)^n |0\rangle 
	&=& |0\rangle + \theta b^{\dagger} a^{\dagger}  |0\rangle - \frac{\theta^2}{2!} abb^{\dagger} a^{\dagger}  |0\rangle 
		                             -  \frac{\theta^3}{3!} b^{\dagger} a^{\dagger} abb^{\dagger} a^{\dagger} |0\rangle 
		                                          \nonumber \\
		 &+& \frac{\theta^4}{4!} abb^{\dagger} a^{\dagger} abb^{\dagger} a^{\dagger} |0\rangle  + \cdots  .
						                	 \label{eq:1333}
\end{eqnarray}
In this expansion,  $\cos \theta_{\mbox{\boldmath $p$}}$ appears in the sum of even-order  terms of $\theta$,  and $\sin \theta_{\mbox{\boldmath $p$}}$ appears  in  the sum of  odd-order terms, and then Eq.(\ref{eq:16}) is yielded.

\section{Metric and energy gap} \label{secA2}
(1) Consider a Bose field $f(x)$  with the following  kinetic energy
\begin{equation}
H_{ef}= -  \int  d^3x    f^{\dagger} (x)  \Delta f(x)   .
  \label{eq:269}
\end{equation}

 The square of the infinitesimal  line-element  is a quadratic function of $dx$ as $dl^2=g_{\mu \nu}(x)dx^{\mu}dx^{\nu}$, where $g_{\mu\nu}(x)$ is a metric tensor  ($\mu,\nu=x,y,z$).
 The inner product  of the field $f(x)$ is defined as
\begin{equation}
  \langle f(x) | f (x) \rangle  =\int \sqrt{|g(x)|} d^3x  f^{\dagger}(x) f (x)  ,
               	\label{eq:16111}
\end{equation}	
where $|g(x)|$ is a determinant of  $g_{\mu\nu}(x)$.  Consider  a gradient of $f$ like $A^{\mu}=g_{\mu\nu}\partial f /\partial x^{\nu} $.  Since the metric depends on the position,  a derivative of a given  vector $A^{\mu}$ with respect  to $x^{\mu}$ is a  covariant derivative as follows 
\begin{equation}
 \frac{DA^{\mu}}{dx^{\mu}}=\frac{dA^{\mu}}{dx^{\mu}}+\Gamma ^{\mu}_{\nu\mu} A^{\nu} ,
               	\label{eq:165}
\end{equation}	
where $\Gamma ^{\mu}_{\nu\mu}$ is a connection coefficient.  This $\Gamma ^{\mu}_{\nu\mu}$  is expressed by  the determinant of the metric tensor $|g(x)|=|g_{\mu\nu}(x)|$ as follows
\begin{equation}
 \Gamma ^{\mu}_{\nu\mu}= \frac{1}{2|g|}\frac{\partial |g|}{\partial x^{\nu}}.
               	\label{eq:17111}
\end{equation}	
With this expression, the covariant derivative of $A^{\mu}$ is given by
\begin{equation}
   \frac{DA^{\mu}}{dx^{\mu}}
     =\frac{1}{\sqrt{|g|}}\frac{\partial(\sqrt{|g|}A^{\mu})} {\partial x^{\mu}} .
               	\label{eq:175}
\end{equation}	

(2)  The  matrix element of the Laplacian operator  is obtained by inserting  $D\partial  /\partial ^2 x$ between $f^{\dagger}(x)$ and $f(x)$ in  Eq.(\ref{eq:16111}).  For $D\partial f /\partial ^2 x$, we  use 
 Eq.(\ref{eq:175}) with $A^{\mu}=g_{\mu\nu}\partial f /\partial x^{\nu}$.  After integration by parts, we get
\begin{eqnarray}
  \langle f(x)| -\Delta |f (x) \rangle   
       &=& -\int \sqrt{|g|} d^3x  f ^{\dagger}\frac{1}{\sqrt{|g|}}
	             \frac{\partial}{\partial x^{\mu}} \left(\sqrt{|g|}g_{\mu\nu}\frac{\partial f}{\partial  x^{\nu}}\right)
	                \nonumber \\
     &=&\int \sqrt{|g(x)|} g_{\mu\nu} d^3x   \frac{\partial f ^{\dagger}}{\partial x^{\mu}}    \frac{\partial f  }{\partial x^{\nu}}  .
               	\label{eq:1801}
\end{eqnarray}	
Our interest  is how  the metric reflecting  permutation symmetry of $f$  looks in the flat space, because  we observe it  as an  effect of Bose statistics in experiments. 
 To rewrite  Eq.(\ref{eq:1801}),  we introduce a new  field $\widehat{f} (x)= |g(x)|^{1/4} f (x)$   as follows
\begin{equation}
  \int \sqrt{|g(x)|}d^3x  f^{\dagger}(x) f(x)=\int d^3x \widehat{f}^{\dagger}(x)\widehat{f}(x)   .
               	\label{eq:185}
\end{equation}	
The gradient in Eq.(\ref{eq:269})  is rewritten  using $ \widehat{f}(x)$  
\begin{equation}
   \frac{\partial f }{\partial x}
   =  |g|^{-1/4} \left(\frac{\partial  }{\partial x}
      -\frac{1}{4}\frac{\partial \ln |g|}{\partial x} \right) \widehat{f}(x)   .
               	\label{eq:19}
\end{equation}	
Hence, the matrix element  of Laplacian in Eq.(\ref{eq:1801})
\begin{equation}
  \langle f (x)| -\Delta | f(x) \rangle 
   =\int d^3x g_{\mu\nu} \left(\frac{\partial  }{\partial x^{\mu}}
        -\frac{1}{4}\frac{\partial \ln |g|}{\partial x^{\mu}} \right) \widehat{f}^{\dagger} (x)
	\left(\frac{\partial  }{\partial x^{\nu}} -\frac{1}{4}\frac{\partial \ln |g|}{\partial x^{\nu}} \right) \widehat{f} (x)   ,
               	\label{eq:195}
\end{equation}	
is rewritten using the integration by parts as follows
\begin{equation}
   \langle f(x)| -\Delta | f (x) \rangle 
    =\int d^3x g_{\mu\nu}\frac{\partial \widehat{f} ^{\dagger} }{\partial x^{\mu}}   \frac{\partial \widehat{f} }{\partial x^{\nu}}
	    + \int W(x) \widehat{f}^{\dagger} (x) \widehat{f} (x) d^3x   ,
               	\label{eq:20001}
\end{equation}	
where
\begin{equation}
  W(x) = \frac{1}{4} \frac{\partial}{\partial x^{\mu}}
                 \left(g_{\mu\nu}\frac{\partial \ln |g|}{\partial   x^{\nu}} \right)
     +\frac{1}{16} g_{\mu\nu} \left(\frac{\partial \ln |g|}{\partial x^{\mu}} \right)
	                   \left(\frac{\partial \ln |g|}{\partial x^{\nu}} \right) .
               	\label{eq:205}
\end{equation}	
This $W(x)$ is a square of the  finite energy gap  in  the excitation spectrum of $ \widehat{f}(x)$  \cite{orl}.

\end{appendices}


\begin{thebibliography}{00}
\bibitem{massless}
 These $u^s(p)$, $v^s(p)$  and $a^s(\mbox{\boldmath $p$})$, $b^{s\dagger }(-\mbox{\boldmath $p$})$  are assumed to be  dimensionless for the later use  ($s$  denotes helicity), and  $V$ is a normalization volume  that gives the mass dimension $3/2$ to $\varphi(x)$.  The dimensionless operators   $a^s(\mbox{\boldmath $p$})$ and $b^{s\dagger }(-\mbox{\boldmath $p$})$ satisfy  $\{a^s(\mbox{\boldmath $p$}), a^{\dagger, s'} (\mbox{\boldmath $p$'}) \} = \{b^s(\mbox{\boldmath $p$}), b^{\dagger, s'} (\mbox{\boldmath $p$'}) \} =\delta_{p,p'}\delta_{s,s'}$, without $\delta(\mbox{\boldmath $p$}-\mbox{\boldmath $p$'})$,
\bibitem{normal}
The real massive particle  recognized by all observers  follows the Lorentz transformation, whereas the massless particle in $|\widetilde{0}\rangle$ is  a fictitious existence.  The Lorentz transformation  does not apply to such an existence,  and therefore this  $|\widetilde{0}\rangle$   is Lorentz invariant in this sense.
\bibitem{bogo}
Equations (\ref{eq:8}) and (\ref{eq:9}) have the same form as the Bogoliubov transformation in superconductivity, but they have different physical meaning in the different situations.    In superconductivity of massive electrons, the Bogoliubov transformation  represents a quasiparticle approximation. In  massless fermions and antifermions in vacuum, Eqs. (\ref{eq:8}) and (\ref{eq:9})   represent not  an approximation,  but   a kinematical relation  that holds exactly.
\bibitem{vec}
 Theoretically,  not only scalar but also  vector mean-field is  possible.
\bibitem{double}
 The  vacuum  $ |\widetilde{0}\rangle$ in Eq.(\ref{eq:16}) does not come from $ {\cal H} _I(x)$, but  from the kinematical reason, and therefore the perturbation of $ {\cal H} _I(x)$ to $L_0^{min}(x)$ in $ |\widetilde{0}\rangle$    does not  doubly  count   the effect of   $ {\cal H} _I(x)$.
\bibitem{massive} 
These   $\widehat{u}^s(p)$ and $\widehat{v}^s(p)$ are the massive-spinor eigen functions, and corresponding operators
    $\widehat{a}(\mbox{\boldmath $p$}) \equiv (2\pi )^{3/2} \widetilde{a}(\mbox{\boldmath $p$})  /\sqrt{V}$  and  $\widehat{b}(\mbox{\boldmath $p$}) \equiv (2\pi )^{3/2} \widetilde{b}(\mbox{\boldmath $p$})  /\sqrt{V}$  
 satisfy  $\{\widehat{a}^s(\mbox{\boldmath $p$}), \widehat{a}^{\dagger, s'} (\mbox{\boldmath $p$'}) \} = \{\widehat{b}^s(\mbox{\boldmath $p$}), \widehat{b}^{\dagger, s'} (\mbox{\boldmath $p$'}) \} =(2\pi)^3 \delta(\mbox{\boldmath $p$}-\mbox{\boldmath $p$'})\delta_{s,s'}$.
\bibitem {gol}
The  Goldstone mode $G$, which appears  in  $\partial_{\mu}G$ of  Eq.(\ref{eq:3851}),  does not contribute to  the non-relativistic limit of scattering amplitude.
\end{thebibliography}
\end{document}